\documentclass{iaus}
\usepackage{graphicx}
\usepackage{natbib}

\renewcommand{\vec}[1]{\mathbf{#1}}

\title[JD 11.~~Perspective acceleration and gravitational redshift on White dwarfs] 
{Perspective acceleration and gravitational redshift. Measuring masses of individual white dwarfs using Gaia + SIM astrometry}

\author[Guillem Anglada-Escud\'e \& John Debes]   
{Guillem Anglada-Escud\'e$^1$
 \and John Debes$^2$}

\affiliation{
$^1$ Dept. of Terrestrial Magnetism,
Carnegie Institution for Science,\\
5241 Broad Branch Rd. Washington DC 20008, USA,\\
email:{\tt anglada@dtm.ciw.edu}
\\[\affilskip]
$^2$ Goddard Space Flight Center, NASA Postdoctoral Program\\
Greenbelt, Maryland 20771, USA\\
email:{\tt john.h.debes@nasa.gov}
}
\pubyear{2009}
\volume{261}  
\pagerange{ -- }
\jname{Relativity in Fundamental Astronomy}
\editors{S. Klioner, P.K.Seidelman \& M.Soffel, eds.}

\begin{document}

\maketitle

\begin{abstract}
According to current plans, the SIM/NASA mission will be launched just after the end of operations for the Gaia/ESA mission. This is a new situation which enables long term astrometric projects that could not be achieved by either mission alone. Using the well-known perspective acceleration effect on astrometric measurements, the true heliocentric radial velocity of a nearby star can be measured with great precision if the time baseline of the astrometric measurements is long enough. Since white dwarfs are compact objects, the gravitational redshift can be quite large ($40$--$80$ km/s), and is the predominant source of any shift in wavelength. The mismatch of the true radial velocity with the spectroscopic shift
thus leads to a direct measure of the Mass--Radius relation for such objects.
Using available catalog information about the known nearby white dwarfs, we estimate how many masses/gravitational redshift measurements can be obtained with an accuracy better than 2\%. Nearby white dwarfs are relatively faint objects (10 $<$ V $<$ 15), which can be easily observed by both missions. We also briefly discuss how the presence of a long period planet can mask the astrometric signal of perspective acceleration.

\keywords{astrometry, (stars:) white dwarfs, relativity, stars: fundamental parameters(masses)}
\end{abstract}

\firstsection 
\section{Introduction}
Perspective acceleration has been previously measured using astrometry for a few nearby stars using ground-based \citep[eg.][]{gatewood:1974} and space-based astrometric observations \citep{dravins:1999}. The motion of a star on the unit sphere can be obtained by the straight forward normalization of the trajectory vector as a function of time.
\begin{eqnarray}
\vec{l}(t) = \frac{\vec{r}_0+\vec{v}\left(t-t_0\right)+\vec{d}\left[t,t_0\right]-\vec{x}_{\rm obs}\left[t\right]}
{\left|\vec{r}_0+\vec{v}\left(t-t_0\right)+\vec{d}\left[t,t_0\right]-\vec{x}_{\rm obs}\left[t\right]\right|}
\end{eqnarray}
\noindent
where the squared brackets $[\ldots]$ indicate explicit dependence on $t$. The vector $\vec{d}$ contains any nonlinear contribution to the stellar motion (eg. a Keplerian orbit). To the second astrometric order, the motion of a star on a locally tangent planet\citep[see][]{anglada:2006} can be written as
\begin{eqnarray}
X_{\rm RA}&=&\left(X_0+\mu_{\rm RA}^*\left(t-t_0\right)+\delta_{\rm  RA}\left[t,t_0\right]-
\pi\,p_{\rm\, RA}\right)\left(1-R\right)\\
Y_{\rm Dec}&=&\left(Y_0+\mu_{\rm Dec}\left(t-t_0\right)+\delta_{\rm  Dec}\left[t,t_0\right]-
\pi\,p_{\rm\, Dec}\right)\left(1-R\right)\\
R&=&R_0+\frac{1}{r_0} v_{\rm r} \left(t-t_0\right)+\delta_{\rm r}\left[t,t_0\right]-
\pi\,p_{\rm\, r}
\end{eqnarray}
\noindent
where $\mu_{\rm RA}^*$ and $\mu_{\rm Dec}$ are the proper motions, the $\delta$'s contain the nonlinear motion terms, $\pi$ is the parallax and $p_{RA}$ \& $p_{Dec}$ are the corresponding parallax factors. The proper motion in $R.A.$ is $\mu_{\rm RA}^* = \mu_{\rm RA} \cos {\rm Dec}$.  This differs from the old convention by the $\cos {\rm Dec}$ term which properly accounts for the deformation of the azimuthal coordinate (R.A) towards the poles. The radial motion is entirely contained in $R$ and contributes as a second order effect. The so-called \textit{perspective acceleration} is the product of the proper motion and the true radial velocity (i.e. $v_r/d$  in $R$) and grows quadratically with time in the direction of the proper motion.

The true radial velocity of the star is obtained if the perspective acceleration can be
observed. Compared to the observed spectroscopic shift, the difference can be attributed to several physical processes, such as the gravitational redshift and the convective outflow. However, the gravitational redshift in white dwarfs ($\sim 70$ km/s) is the dominant source of non-kinematic shifts. The astrometric signal of the offset induced by the \textit{gravitational redshift} for a white dwarf is
\begin{eqnarray}
\alpha_{\rm GR} = \mu \frac{1}{r_0}v_{\rm GR}\left(t-t_0\right)^2\,;\,\,\,
v_{\rm GR} = \frac{GM}{Rc} \sim 68.9 \,{\rm km/s}\, \frac{M_*}{M_{\rm sun}}\frac{R_{\rm earth}}{R_{*}}
\end{eqnarray}
\noindent
 This method was first demonstrated by \citet{gatewood:1974} on the Van Maneen 2 star. The gravitational redshift has also already been measured in binary systems such as Sirius B. Gaia+SIM offers the opportunity of applying the method to a statistically significant number of nearby white dwarfs. White dwarfs can be used as absolute stellar candles to obtain distance measures and absolute photometry to old stellar complexes such as globular clusters and nearby galaxies \citep{davis:2008}.
\section{Perspective acceleration in the presence of planetary companions}
\begin{figure}[bt]
\begin{center}
 \includegraphics[width=4.2in, clip]{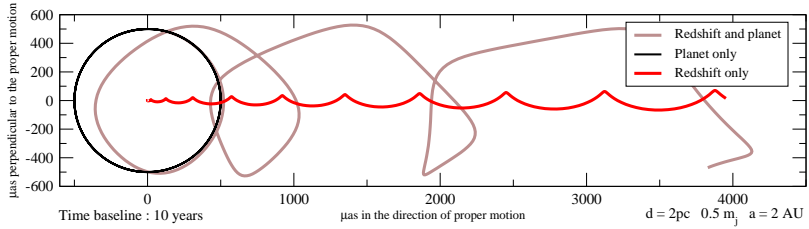}
 \caption{Apparent motion of a star during 10 years due to a
 planetary companion(black), perspective acceleration(red),
 and the combined effect(brown). Both axis are in $\mu$as.}
   \label{fig2}
\end{center}
\end{figure}

Long term astrometric observations will be extremely sensitive to planets around white dwarfs. Most planetary companions to white dwarfs are expected to lie in regions exterior to a primordial separation of $3$-$5$ AU \citep{villaver:2007}, which due to central star mass loss on the red giant branch corresponds to final separations of between $6$--$12$ AU or larger. The presence of a long period planet (P$>$15 years) should be obvious when combining Gaia+SIM data sets, but its signature will be strongly correlated with the perspective acceleration effect. The combined signal of a planet with a relatively short period compared to the time-span of the observations is shown in Fig.~\ref{fig2}. In the short period case the decoupling of the perspective acceleration with the planet signal is feasible.
In the case of a long period planet, only the orbital motion perpendicular to the proper motion can be measurable independently. While these systems are interesting from the point of view of planetary systems evolution, they are not useful for gravitational redshift studies.

\section{The nearby white dwarf sample}
We have collected a sample of white dwarfs in the Solar neighborhood (d$<$20 pc). We compiled the sample from \citet{holberg:2008A} with proper motions obtained from SIMBAD. In our knowledge, it represents the most complete sample of WD within $20$ pc. The distances determined are trigonometric parallaxes if available. Except for a few binaries (eg. Sirius B), the masses are based on model estimates \citet{holberg:2008B}. The gravitational redshift and its astrometric signal
have been simulated in Gaia-alone (5 years), and Gaia+SIM (5 years + 4 SIM epochs at $10\mu$as), obtaining a realistic estimate of what can be obtained for each star.

\section{Conclusions}
We show how the gravitational redshift term is very significant in the apparent motion of the star and can be measured by comparing the perspective acceleration effect(astrometry) and the Doppler shift (spectroscopy).
\begin{figure}[tb]
\begin{center}
 \includegraphics[width=3.5in]{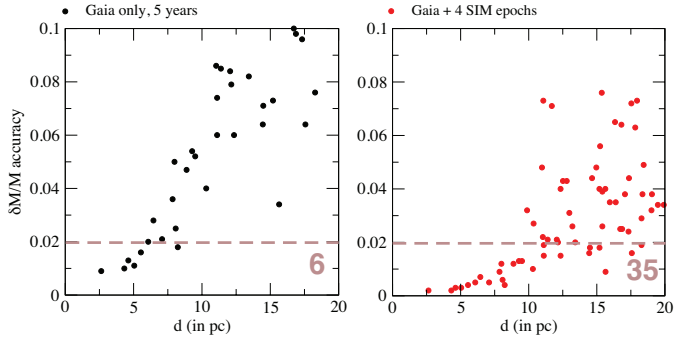}
 \caption{Each point represents a nearby White Dwarf. The relative
 accuracy at which the mass can be determined is plotted against
 the distance. The left panel shows Gaia observations alone(5 years). The right
 panel shows Gaia+SIM(10 years).}
   \label{fig3}
\end{center}
\end{figure}
We find that combining Gaia+SIM observations, the gravitational redshifts of 35 already known objects can be measured with an accuracy of 2\%. This number may grow as new nearby white dwarf candidates are identified. The redshift of ALL the known white dwarfs within 20 pc can be obtained with an accuracy better than 10\% with 4 additional SIM epochs added to the Gaia ones.

\textbf{Acknowledgments.} This work is part of the SIM Science study : Gaia-SIM Legacy project funded by JPL/Caltech under NASA contract NMO 710776, which explores science cases benefiting from long term astrometric observations combining Gaia and SIM.

\end{document}